\documentclass[aps,pra,showpacs,showkeys,floatfix,nofootinbib,superscriptaddress]{revtex4-1}
\usepackage{amssymb,amsmath,amsfonts,amsthm,amscd}
\usepackage[utf8]{inputenc}
\usepackage{graphicx}
\usepackage{verbatim}
\usepackage{tensor}
\usepackage{hyperref}
\usepackage[all,cmtip]{xy}

\hypersetup{
  pdftitle={Categorical Quantum Circuits},
  pdfauthor={Bergholm, Ville and Biamonte, Jacob D.},
  pdfsubject={category theory, quantum circuits},
  pdfkeywords={category theory, quantum circuit model, quantum information, tensor network, categorical quantum mechanics}
}

\newcommand{\QC}{{\sf{QC}}}

\newcommand{\bbone}{\openone}
\newcommand{\bra}[1]{\mbox{$\langle #1|$}}
\newcommand{\ket}[1]{\mbox{$|#1\rangle$}}
\newcommand{\braket}[2]{\mbox{$\langle #1|#2\rangle$}}
\newcommand{\ketbra}[2]{\mbox{$|#1\rangle\langle #2|$}}

\DeclareMathOperator{\Tr}{Tr}

\newcommand{\eg}{e.g.}
\newcommand{\ie}{i.e.}

\newcommand{\gate}[1]{\ensuremath{\text{\sc #1}}}
\newcommand{\ADD}{\gate{ADD}}
\newcommand{\NADD}{\gate{NADD}}
\newcommand{\SWAP}{\gate{SWAP}}
\newcommand{\CNOT}{\gate{CNOT}}
\newcommand{\NOT}{\gate{NOT}}
\newcommand{\NEG}{\gate{NEG}} 

\newcommand{\opstate}[1]{\left \ulcorner #1 \right \urcorner}

\newcommand{\COPY}[2]{\gate{COPY}\ensuremath{_{#1}^{#2}}}
\newcommand{\PLUS}[2]{\gate{PLUS}\ensuremath{_{#1}^{#2}}}
\newcommand{\DOT}[2]{\gate{D}\ensuremath{_{#1}^{#2}}}
\newcommand{\prune}{\star}

\newcommand{\I}{\bbone} 
\newcommand{\eye}{\mathbf{1}} 

\newcommand{\Figref}[1]{Fig.~\ref{#1}}
\newcommand{\Eqref}[1]{Eq.~\eqref{#1}}
\newcommand{\be}{\begin{equation}}
\newcommand{\ee}{\end{equation}}

\newcommand{\cat}[1]{\ensuremath{\mathcal{#1}}}
\newcommand{\Hom}[3][C]{\cat{#1}(#2,#3)}
\newcommand{\Ob}[1]{\text{ob}(\cat{#1})}
\newcommand{\F}[1]{F}  

\newcommand{\hilbert}[1]{\ensuremath{\mathcal{#1}}}

\newcommand{\cupket}[1]{\ket{\cup}_{#1 \otimes #1}}
\newcommand{\capbra}[1]{\bra{\cup}_{#1 \otimes #1}}

\newcommand{\isom}{\cong}  

\theoremstyle{plain}
\newtheorem{theorem}{Theorem}

\newtheorem{corollary}[theorem]{Corollary}

\theoremstyle{definition}
\newtheorem{remark}[theorem]{Remark}
\newtheorem{definition}[theorem]{Definition}
\newtheorem{example}[theorem]{Example}

\newcommand{\HH}{\hilbert{H}}
\newcommand{\C}{{\mathbb C}}
\newcommand{\K}{{\mathbb K}}

\begin{document}
\title{Categorical Quantum Circuits}
\author{Ville Bergholm}
\email{ville.bergholm@iki.fi}
\affiliation{Department of Applied Physics, Aalto University School of
  Science and Technology, Espoo, Finland}

\author{Jacob D. Biamonte}
\email{jake@qubit.org}
\affiliation{Oxford University Computing Laboratory, Oxford, UK}

\pacs{03.65.Fd, 03.65.Ca, 03.65.Aa}
\keywords{category theory, quantum circuit model, quantum information, tensor network}

\begin{abstract}
In this paper, we extend past work done on the application of the
mathematics of category theory to quantum information science.
Specifically, we present a realization of a dagger-compact category
that can model finite-dimensional quantum systems
and explicitly allows for the interaction of systems of
arbitrary, possibly unequal, dimensions. Hence our framework
can handle generic tensor network states, including matrix product
states.
Our categorical model subsumes the traditional quantum circuit model
while remaining directly and easily applicable to problems stated in the language
of quantum information science.
The circuit diagrams themselves now become
morphisms in a category, making quantum circuits a special case
of a much more general mathematical framework.
We introduce the key algebraic properties of our tensor calculus
diagrammatically and show how they can be applied to solve
problems in the field of quantum information.

\end{abstract}
\maketitle


\section{Introduction}


Diagrammatic methods in physics and in quantum information science
have a long history~\cite{yao93,BDEJ95,ElementaryGates:95,KSV02}.
Their importance stems from the fact that they
enable one to perform mathematical reasoning and even actual calculations using
intuitive graphical objects instead of abstract mathematical entities.
Modern quantum circuit diagrams (QCDs) are highly sophisticated tools,
even though many of their features were developed in a largely \emph{ad hoc}
manner.

There has been recent interest in the application of tensor calculus
and related diagrammatic methods to problems in condensed matter physics.
In the present paper, we develop a generalization of the methods of
categorical quantum mechanics~\cite{catQM} to the case where the systems are of
arbitrary dimensions.  This generalization is a logical next step to
extend the existing categorical framework, and is one of practical
importance.
For instance, in matrix product states, the so called
bond dimension will in general vary between three-index tensors
attached to a physical spin, and so a full categorical treatment must
be able to handle the varying dimension of the internal legs.
The approach we have taken is to define a dagger-compact
category~\QC, which takes this scenario into account explicitly.
Our work is the first to
consider general multi-valued quantum networks and to enable one to deal with
the full class of tensor network states (and hence the abovementioned
subclass of matrix product states) categorically.
The work in~\cite{biamonte2010} was the first to consider the application of category theory in tensor network states.  
There several results are proven related to tensor network theory, with the dimension of all tensor legs fixed to some arbitrary constant.

Apart from general tensor networks, our results also apply to quantum
circuits.  In particular, we are able to build on ideas appearing in
the gate model of quantum computing and incorporate them here.
We use this framework to define a calculus of $d$-dimensional quantum
logic gates, and recover the standard qubit results when~$d=2$.
The calculus necessitates the introduction of a negation gate which reduces
to the identity operator in the qubit case, illustrating its special nature.
We also define a class of tensors which we call symmetric dots.
They can be used \eg{} to factor certain multi-qudit gates, including
the generalized \CNOT{} gate,
into a network of dots and single-qudit gates.

Category theory is often used as a unifying language
for mathematics, and in more recent times to formulate physical
theories~\cite{baez2009,prehistory}.  One of the strong points of the area of
applied mathematics known as categorical modeling is that it comes equipped with a powerful
graphical language that can be proven to be fully equivalent to the
corresponding algebraic notation. We use this to define the algebraic
properties of the network components we consider here
diagrammatically.
Category theory has only recently been used to model quantum mechanics~\cite{catQM}. 
Connecting categorical methods to the established area of
quantum circuit theory and tensor network states seems reasonable as
category theory provides the
exact arena of mathematics concerned with the diagrammatic reasoning
present in the existing methods to manipulate quantum networks.  
These \emph{string diagrams} capture the mathematical properties of
how maps (states and operators in the circuit model) can be composed.
By considering the categorical description of the mathematics used
in quantum mechanics, one essentially gets quantum circuits for free! 

Traditional QCDs are graphs that are
planar, directed and acyclic. These are a subclass of the graphs one
can construct in a dagger-compact category.  We take steps beyond this by
considering the symmetric compact structure of the category.  This amounts to
adding in maps that are equivalent to Bell states and Bell
effects: one can then arrive \eg{} at the well known results
surrounding channel-state duality.  However, categorical
dualities come with something novel: an intuitive graphical
interpretation which we use to manipulate quantum circuits in new ways and which
exposes channel-state duality as a consequence of simply bending wires.  For
instance, by temporarily
dropping causality (directed temporal ordering) one can with relative ease
perform very nontrivial operations on the diagrams and then convert them back into a
standard, physically implementable quantum circuits.

Both category theory and the quantum circuit model are well developed fields,
backed by years of fundamental research.  The state of the art in graphical
languages used in quantum information science can be found \eg{} in~\cite{GESP07}.
We also note the work in~\cite{Atemp06}.
As in the theory of tensor network states~\cite{GESP07}, the theory of
categories allows one to study the mathematical
structure formed by the composition of processes themselves (see for instance work on
tensor network states~\cite{biamonte2010}).

Our main focus in this article is to connect these two fields: the mathematical
ideas appearing in category theory with the state of the art in quantum
information science. By showing how the structure of a dagger-compact
category can be represented in a quantum circuit, and by showing how
quantum circuits can be transformed using methods from category theory
we aim to derive results useful to both areas.

\subsection*{Background reading}

This work attempts to be mostly self contained. For those interested
in the string diagrams,
Selinger's ``A survey of graphical languages for monoidal categories''
offers an excellent starting
place~\cite{Selinger09}.  The mathematical insight behind using pictures to
represent these and related networks dates back to Penrose and in quantum
circuits, to Deutsch.  The mathematics behind category theory
is based largely on a completeness result (originally proved by Joyal and
Street) about the kinds of string diagrams we consider here~\cite{JS91, MacLane, Selinger09}.

We build on ideas across several fields, see e.g. the course notes \cite{course}.  This includes the work
by Lafont~\cite{boolean03} which was aimed at providing an algebraic
theory for classical Boolean circuits and in particular \cite{Lafont92} (which is available free online).  
The first application of categories in quantum computing seem to be found in \cite{boolean03, catQM}.  The work~\cite{catQM} considered a categorical model of quantum protocols, and produced a framework called ``categorical quantum mechanics''.  By considering the composition of algebraically defined building blocks, the work~\cite{biamonte2010}
put forth the building blocks needed for a categorical theory of tensor network states.  By considering quantum observables, the work \cite{redgreen} recently took a different direction to the one we explore here and made progress towards a categorical model of quantum theory that could be applicable to problems in quantum information and computing.  Recently several tutorials on categories and the corresponding diagrammatic calculus have been made available.  Our favorites include~\cite{baez2009, Selinger09}.

\section{Extended quantum circuit diagrams}

We will now begin our presentation of an extended form of the existing diagrammatic
notation for describing quantum circuits.
Some of these concepts were first introduced
in~\cite{catQM, catprotocols, redgreen}, where the
authors derived a categorical
representation that was expressive enough to represent many of the
components used in standard quantum circuits.  In a seemingly independent research
track, some of these concepts also appeared in~\cite{GESP07,Atemp06} as well as related work.  

Each extended quantum circuit diagram corresponds to a
single morphism in the category~\QC.
The main difference to ordinary QCDs is that
an extended diagram \emph{does not have to correspond to a quantum operation.}
One of the key benefits of these
diagrams is that they can be manipulated in a very intuitive, visual way.
Objects (boxes etc.) on wires can be slid along them.
The wires themselves can be bent and rearranged.
Nodes where several wires meet can often be combined
and split according to simple rules. After such changes, the
diagram can be converted back into an ordinary, physically implementable quantum
circuit, depending on the specific application.

\begin{definition}[The category of quantum circuits \QC]
\label{def:QC}
\QC{} is a category that consists of
\begin{enumerate}
\item Objects $A := (\hilbert{A}, D_A)$, where
$D_A = ({d_A}_i)_{i=1}^{n_A}$ is a list of positive integer dimensions
and
$\hilbert{A} = \C^{d_1} \otimes \C^{d_2} \otimes \cdots \otimes \C^{d_{n_A}}$
is a finite dimensional complex Hilbert space.
The dimension of~$A$ is $\dim A := \dim \hilbert{A} = \prod_{i=1}^{n_A} {d_A}_i$.
$n_A$~denotes the number of \emph{subsystems} in the object.
If $n_A = 1$ the object is \emph{simple},
otherwise it is \emph{composite}.

For the Hilbert space in each object~$A$ we
shall choose a \emph{computational basis} (equal to the
standard tensor basis), denoted
by~$\{ \ket{i_1 i_2 \ldots i_{n_A}}_A \}_{i_k=0}^{{d_A}_k-1}$ and ordered
in the big-endian fashion.

\item For every pair of objects $A, B$ the set of morphisms $\Hom[\QC]{A}{B}$,
which consists of all bounded linear maps
between the Hilbert spaces $\hilbert{A}$ and $\hilbert{B}$.
$D_A$ and $D_B$ are the input and output dimensions, respectively, of
the morphisms in this set.
A unitary \QC-morphism is also called a \emph{gate}.

\item Composition of morphisms~$\circ$, which is just the usual composition of
  linear maps.

\item Tensor product bifunctor~$\otimes$ with the unit object
$\eye := (\C, (1))$. The tensor product of objects is given by
$A \otimes B := (\hilbert{A} \otimes \hilbert{B}, D_A \star D_B)$,
where
$\star$~denotes list concatenation with the elimination of
unnecessary singleton dimensions.
The tensor product of morphisms 
 $f \otimes g$, where $f: A \to A', \: g: B \to B'$, is given by the
Kronecker product of the corresponding matrices in the computational basis:

\[
\bra{ij}_{A' \otimes B'} (f \otimes g) \ket{pq}_{A \otimes B} := \bra{i}_{A'} f \ket{p}_A \bra{j}_{B'} g \ket{q}_B
\]

\item Dagger functor~$\dagger$, which is identity on the objects and
  takes the Hermitian adjoint of the morphisms.

\item
  For every object~$A$ the unit and counit morphisms, defined in terms of the
  computational basis:
\begin{align*}
\eta_A := \sum_k \ket{kk}_{A \otimes A}, \qquad
\epsilon_A &:= \sum_k \bra{kk}_{A \otimes A} = \eta_A^\dagger.
\end{align*}
Every object is its own dual: $A^* = A$.

\end{enumerate}

\QC{} is categorically equivalent to a strictified~\cite{MacLane} {\sf FdHilb} with
explicit dimensional typing.
\end{definition}

\begin{theorem}
\QC{} is a dagger-compact category.
\end{theorem}

\begin{proof}[Proof sketch]~
\begin{itemize}
\item
It is straightforward to show that \QC{} is a category: composition of
the morphisms is clearly
associative, and for each object~$A$ the corresponding identity
morphism~$1_A$ is the identity map $\I_\hilbert{A}$ on $\hilbert{A}$.

\item
\QC{} is also seen to be monoidal.
The tensor product
fulfills the covariant bifunctor rule
\[
(g \circ f) \otimes (t \circ s) = (g \otimes t) \circ (f \otimes s).
\]
It is associative,
$(A \otimes B) \otimes C \isom A \otimes (B \otimes C)$,
and $\eye \otimes A \isom A \otimes \eye \isom A$.
The associator $\alpha_{ABC}$ and the left and right unitor isomorphisms
$\lambda_{A}$ and $\rho_{A}$ are trivial and fulfill the
the pentagon and triangle axioms.

\item
The symmetric braiding isomorphism
$c_{A,B}: A \otimes B \to B \otimes A$,
required to make \QC{} symmetric monoidal,
is given by the \SWAP{} gate:
$c_{A,B} := \SWAP_{A,B} := \sum_{ab} \ket{ba}_{B \otimes A}
\bra{ab}_{A \otimes B} = c_{B,A}^{-1}$
which fulfills the hexagon axiom.

\item
The dagger is a contravariant endofunctor, 
is easily seen to be involutive ($f^{\dagger \dagger} = f$) and has the proper
interaction with the composition and tensor product:
$(g \circ f)^\dagger = f^\dagger \circ g^\dagger$, and
$(f \otimes g)^\dagger = f^\dagger \otimes g^\dagger$.
Furthermore, the isomorphisms $\alpha$, $\lambda$, $\rho$ and $c$ are all unitary
so \QC{} is dagger symmetric monoidal.

\item
The unit and counit morphisms fulfill the adjunction triangles (``snake equations''),
and are symmetric, and thus
\be
\label{eq:unitsymmetry}
c_{A, A^*} \circ \epsilon_A^\dagger = c_{A, A} \circ \eta_A^{\dagger\dagger} =
c_{A, A} \circ \eta_A = \eta_A.
\ee
Together with the other properties this makes \QC{} a dagger-compact
category.
\end{itemize}
\end{proof}

\begin{remark}[Quantum circuits as \sf{PROP}s]  
Our definition of \QC{} is motivated by the fact that
quantum computations
typically take place in a fixed Hilbert space with each subsystem labeled.
Symmetric monoidal categories with fixed types (all the objects are,
say, qubits) are called {\sf{PROP}s}~\cite{PROPs}.
We do not use this construction since we want to be able to handle
also systems composed of different types of subsystems, \eg{} qubits \emph{and} qutrits.
\end{remark}

\begin{remark}[Basic notational differences to standard string diagrams]
To make our presentation more approachable to readers who
have a background in quantum information science (as opposed to
category theory), we have decided to depart from certain common
category theory conventions.

\begin{itemize}
\item
As in standard QCDs, in the present diagrams time flows from left to
right across the page. This is in contrast to the diagrams in some category theory
texts in which time flows either downwards or upwards.
Likewise,
in the diagram of $A \otimes B$ the line representing~$A$ is drawn above
the line representing~$B$ and not the other way around.

\item
In addition to the usual unit and counit morphisms $\eta$ and $\epsilon$
we define the corresponding normalized states and costates:
$\cupket{A} = \frac{1}{\sqrt{d}} \eta_A$ and
$\capbra{A} = \frac{1}{\sqrt{d}} \epsilon_A$,
where $d = \dim A$.
They correspond to physically realizable entangled
states and help to keep the normalization of the diagrams explicit.

\item
We do not use dual spaces in implementing the compact structures 
but rather choose a preferred \emph{computational basis}, as is common
in quantum computing. Hence our wires do not have directional markers
on them.
\end{itemize}
\end{remark}

\subsection{Basic definitions}


\begin{remark}[Einstein summation convention]
We make use of Einstein notation for covariant
and contravariant tensor indices, along with the usual summation
convention (indices appearing once as a subscript and once as a
superscript in the same term are summed over)
whenever the summation limits are evident from the context.

\end{remark}

\begin{definition}[Computational basis]
\label{def:compbasis}
For every $d$-dimensional Hilbert space~\hilbert{H} we
shall choose a computational basis (also called the $z$-basis),
as explained in Def.~\ref{def:QC}.
We use modulo~$d$ arithmetic for the basis vector indices, using the
symbols $\oplus$ and $\ominus$ to denote modular addition and subtraction.
For composite objects, the index needs to be converted to a linear
index first in the big-endian fashion.

When necessary, a subscript
outside an operator, a ket or a bra is used to denote the system it
acts on or corresponds to.
\end{definition}

\begin{definition}[Discrete Fourier transform gate]
We denote the discrete Fourier transform gate by~$H$:
\be
\label{eq:H}
H_{\hilbert{H}} := \frac{1}{\sqrt{d}} \sum_{ab} e^{i 2 \pi a b/d}
\ketbra{a}{b}_{\hilbert{H}},
\ee
where $d = \dim \hilbert{H}$ is the dimension of the Hilbert space the gate
acts in.
We can see that $H^T = H$, and that
in a qubit system $H$~coincides with the one-qubit Hadamard gate.
\end{definition}

\begin{definition}[$x$-basis]
Essentially, the discrete Fourier transform~$H_{\hilbert{H}}$ is a transformation
between two mutually unbiased bases, the computational basis and the
\emph{$x$-basis} $\{\ket{x_k}_{\hilbert{H}}\}_{k=0}^{d-1}$, defined as
\be
\ket{x_k}_{\hilbert{H}} := H \ket{k}_{\hilbert{H}}.
\ee
\end{definition}

\begin{figure}[t]
\includegraphics[width=0.6\textwidth]{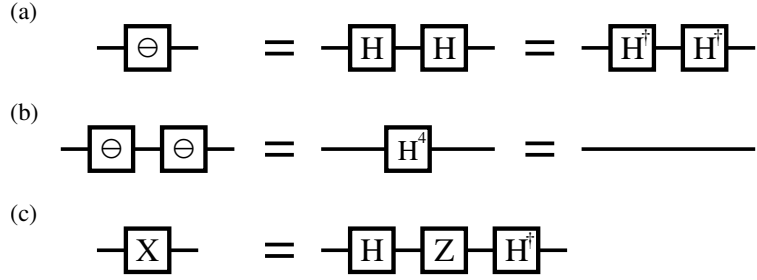}
\caption{Basic utility gates.
(a)~The \NEG{} gate performs a modular negation in the computational basis:
 $\NEG: \ket{k} \mapsto \ket{\ominus k}$.
 It can be implemented using the discrete Fourier transform gate~$H$.
(b)~Negation gate squared equals identity: $\NEG^2 = H^4 = \I$.
(c)~The $X$~gate modularly increments by one in the computational
  basis: $X: \ket{k} \mapsto \ket{k \oplus 1}$. The $Z$~gate does the same in the $x$-basis.
}
\label{fig:qftneg}
\end{figure}

\begin{definition}[Negation gate]
The negation gate is defined as
\be
\NEG_{\hilbert{H}} := H_{\hilbert{H}}^2 = H_{\hilbert{H}}^{\dagger 2} = \sum_a \ketbra{\ominus a}{a}_{\hilbert{H}}.
\ee
As the name suggests, it performs a negation modulo~$d$ in the
computational basis. As one would expect we have
$\NEG^2 = H^4 = \I$, as shown in \Figref{fig:qftneg}.
In a qubit system the negation gate reduces to the identity operator.
\end{definition}

\begin{definition}[Generalized $Z$ and $X$ gates]
We define the generalized $Z$ and
$X$ gates
in the $d$-dimensional Hilbert space~$\hilbert{H}$
as follows~\cite{PhysRevA.68.022310}:
\begin{align}
\label{eq:ZX}
Z_{\hilbert{H}} &:= \sum_k e^{i 2 \pi k/d} \ketbra{k}{k}_{\hilbert{H}},\\
X_{\hilbert{H}} &:= \sum_k \ketbra{k \oplus 1}{k}_{\hilbert{H}}.
\end{align}
In fact, the $Z$ and~$X$ gates are the same operator in two different
bases, related through conjugation with~$H$:
\begin{align}
H X H^\dagger &= Z,\\
H Z H^\dagger &= X^\dagger.
\end{align}
$X$ has the $x$-basis as its eigenbasis and modularly increments a computational basis state by~$1$,
whereas $Z$~is diagonal in the computational basis and modularly increments
$x$-basis states:
\begin{align}
\notag
X^a \ket{k} &= \ket{k \oplus a},\\
Z^a \ket{x_k} &=
H X^a H^\dagger H \ket{k} =
H \ket{k \oplus a} =
\ket{x_{k \oplus a}}.
\end{align}
Consequently we have~$Z^d = X^d = \I$.
Furthermore,
\begin{align}
Z^a X^b &= e^{i 2 \pi a b/d} X^b Z^a \quad \text{and}\\
\Tr(Z^a X^b) &= d \, \delta_{a,0} \, \delta_{b,0}.
\label{eq:ZXtrace}
\end{align}
\Figref{fig:qftneg} presents the gate symbols we use for the $Z$ and
$X$ gates.
When \hilbert{H} is a qubit, $X$ and $Z$ reduce to the Pauli
matrices $\sigma_x$ and $\sigma_z$, respectively.
\end{definition}

\begin{figure}[t]
\includegraphics[width=0.6\textwidth]{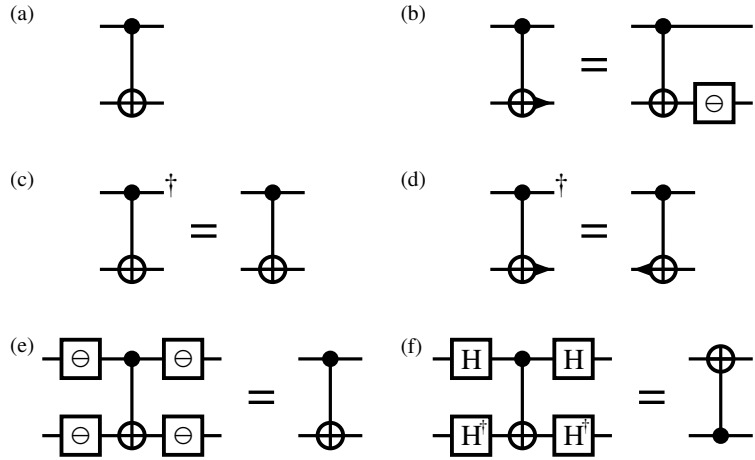}
\caption{
(a)~\NADD{} gate: $\ket{x,y} \mapsto \ket{x,\ominus x \ominus y}$.
(b)~\ADD{} gate: $\ket{x,y} \mapsto \ket{x, x \oplus y}$.
(c,d)~\NADD{} is self-inverse, \ADD{} is not, hence the need for the
arrow-like symbol denoting the output direction.
(e,f)~Identities involving \NADD{}, \NEG{} and~$H$.}
\label{fig:add}
\end{figure}

\begin{definition}[Modular adder gate]
We define the modular adder gate \ADD{} as
\be
\label{eq:add}
\ADD_{i,j} := \sum_{xy}  \ketbra{x,y \oplus x}{x,y}_{i,j}.
\ee
The negated modular adder gate, \NADD, is obtained by negating the
``result qudit'' of the \ADD{} gate:
\be
\label{eq:nadd}
\NADD_{i,j} := \NEG_{j} \ADD_{i,j} = \sum_{xy}  \ketbra{x,\ominus x \ominus y}{x,y}_{i,j}.
\ee
In a two qubit system, both \ADD{} and \NADD{} reduce to the
\gate{CNOT} gate, and thus can be understood as its
higher-dimensional generalizations.
\NADD{} is self-inverse while \ADD{} is not, which is why we choose to
use the traditional \gate{CNOT} symbol to denote \NADD{} in general.
For \ADD{} we add a small arrow to denote the
output direction.
\Figref{fig:add} presents the gate symbol and
identities involving \ADD, \NADD{} and~$H$.
\end{definition}

\begin{definition}[Generalized plus state] 
We define the generalized $\ket{+}_{\hilbert{H}}$ state as
\be
\ket{+}_{\hilbert{H}} := \ket{x_0}_{\hilbert{H}} = H \ket{0}_{\hilbert{H}} = \frac{1}{\sqrt{d}}
\sum_{i=0}^{d-1} \ket{i}_{\hilbert{H}},
\quad \text{where} \; d = \dim \hilbert{H}.
\ee
\end{definition}

\begin{definition}[Generalized Bell states]
The concept of a Bell state, normally defined for two-qubit systems,
can be generalized to systems of two $d$-dimensional qudits.
In this case the Bell states $\{\ket{B_{a,b}}_{\hilbert{H} \otimes \hilbert{H}}\}_{a,b=0}^{d-1}$
are a set of $d^2$~orthonormal and maximally entangled
two-qudit states. They are parameterized by two integers, $a$ and~$b$,
and can be prepared using the Fourier and \ADD{} gates:
\be
\ket{B_{a,b}}_{i,j}
:= 
\frac{1}{\sqrt{d}} \sum_k e^{i 2 \pi a k/d} \ket{k, k \oplus b}_{i,j}
= \ADD_{i,j} H_i \ket{a,b}_{i,j}.
\ee
\end{definition}

We will now proceed to describe the circuit elements appearing in the
extended QCDs.

\subsection{Systems as \QC-objects}

\begin{figure}[h]
\includegraphics[width=0.25\textwidth]{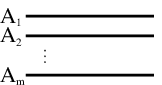}
\caption{Composite \QC-object~$A = A_1 \otimes A_2 \otimes \ldots \otimes A_m$,
corresponding to an $m$-partite system with the Hilbert space
$\hilbert{A} \isom \hilbert{A}_1 \otimes \hilbert{A}_2 \otimes \ldots \otimes \hilbert{A}_m$.
}
\label{fig:wires}
\end{figure}

In our diagrams, much like in ordinary QCDs, time
flows from left to right.\footnote{In converting a diagram into an algebraic expression one
  needs to reverse the left-right order since traditional quantum
  mechanics uses left multiplication to represent operations on states.}
Horizontal wires each describe individual
quantum systems (simple \QC-objects). Stacking the wires vertically
corresponds to a system comprised of several subsystems
(a composite \QC-object), as shown in~\Figref{fig:wires}.
Unless the types are clear from the context, each
wire should be explicitly labeled.

Alternatively, a wire $A$ can be understood as the identity morphism~$1_A$.
The unit object for the tensor product, $\eye$, is represented by
empty space.

Unlike in standard QCDs, the wires are allowed to deviate from a
straight horizontal line and even cross each other (which corresponds
to swapping the order of the corresponding subsystems using the
symmetric braiding isomorphism~$c$), as long as
they remain \emph{progressive} from left to right, and the
relative order of the endpoints of different wires does not change.
As we shall see in Sec.~\ref{sec:cups}, a wire reversing its direction of progression
has a specific meaning.

\subsection{Morphisms: states and operators on equivalent footing}
Category theory allows one to study the mathematical
structure formed not only by the composition of processes but 
also the composition of states. This becomes evident once we define both
states and operators as morphisms in the category~\QC. In the diagrams,
the morphisms are represented by geometrical shapes connected to the
wires.
The only exceptions to this rule are the identity morphisms
(represented by the wires themselves), and
morphisms of the type
$f:\eye \to \eye$, also called \emph{scalars}.
Since the tensor unit object~$\eye$ is represented
by empty space and $f$~commutes with all morphisms, its representation
is just the number~$f(1)$ anywhere in the diagram.

\subsubsection{States as \QC-morphisms}\label{sec:QSmorphisms}

In \QC, a pure state $\ket{\psi}$
represented by a ket, or a ray in a Hilbert space~$\hilbert{A}$,
corresponds to a linear map from $\C$ to~$\hilbert{A}$, or the morphism
\begin{equation} 
\psi: \eye \to A, \quad z \mapsto z \ket{\psi}.
\end{equation}
For instance, consider the two-qubit state
$\ket{\Psi_+} = \frac{1}{\sqrt{2}}(\ket{01}+\ket{10})$: this corresponds to the map
$\C \xrightarrow{~\Psi_+~} \C^2 \otimes \C^2$ in the category~\QC.

\begin{figure}[h]
\includegraphics[width=0.2\textwidth]{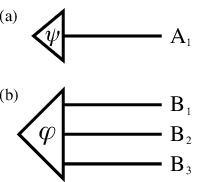}
\caption{
(a) State $\ket{\psi}$ with a single subsystem.
(b) State $\ket{\varphi}$ with three subsystems.
$\psi$ is a morphism of type $\eye \to A_1$ and 
$\varphi$ is a morphism of type $\eye \to B_1 \otimes
B_2 \otimes B_3$. }
\label{fig:states}
\end{figure}

In a diagram, a pure state (or equivalently the corresponding state
preparation procedure)
is represented by a left-pointing labeled
triangle with a number of wires extending from its base to the right,
as shown in \Figref{fig:states}.
Each wire corresponds to a subsystem of the state.
Flipping a triangle horizontally converts it into the corresponding
costate (bra), and can be understood as a projective measurement with
postselection (an \emph{effect}).

A state $\ket{\psi}$ can be expanded in the computational
basis,
resulting in the presentation
\be
\ket{\psi}_{A} = \psi^{a_1 \cdots a_m} \ket{a_1 \cdots a_m}_{A_1 \otimes A_2 \otimes \ldots \otimes A_m}.
\ee

\subsubsection{Operators as \QC-morphisms}

Operators, or bounded linear maps from one Hilbert
space to another, can be identified with the morphisms
in \QC{}. As an example we can consider quantum gates, unitary maps from
a Hilbert space to itself.

\begin{figure}[h]
\includegraphics[width=0.3\textwidth]{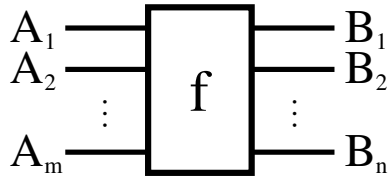}
\caption{Morphism $f: A \to B$, where $A$ and $B$ have $m$ and $n$
  subsystems, respectively.}
\label{fig:ops}
\end{figure}

In the diagrams, operators are represented using labeled boxes on the wires,
as shown in \Figref{fig:ops}.
Assume that we have a morphism $f: A \to B$, and that
the domain and codomain \QC-objects are
tensor products of simple \QC-objects
given by
$A = A_1 \otimes A_2 \otimes \ldots \otimes A_m$ and
$B = B_1 \otimes B_2 \otimes \ldots \otimes B_n$.
This means that the diagram for $f$ has $m$~input legs and $n$~output legs.
For certain operators, such as the
\ADD{} gate, we introduce specific symbols.
 
Using the computational basis
we can present~$f$ as
\be
f = \ketbra{b_1 \cdots b_n}{b_1 \cdots b_n}_{B} f \ketbra{a_1 \cdots a_m}{a_1
\cdots a_m}_{A}
= \ket{b_1 \cdots b_n}_{B} f\indices{^{b_1 \cdots b_n}_{a_1 \cdots a_m}}
\bra{a_1 \cdots a_m}_{A}.
\ee
Given a state $\ket{\psi}_{A} = \psi^{a_1 \cdots a_m} \ket{a_1 \cdots a_m}_{A}$,
we have
$f \ket{\psi}_{A} =
f\indices{^{b_1 \cdots b_n}_{a_1 \cdots a_m}} \psi^{a_1 \cdots a_m}
\ket{b_1 \cdots b_n}_{B}$.

\subsubsection{Composition and tensor product}

\begin{figure}[h]
\includegraphics[width=0.85\textwidth]{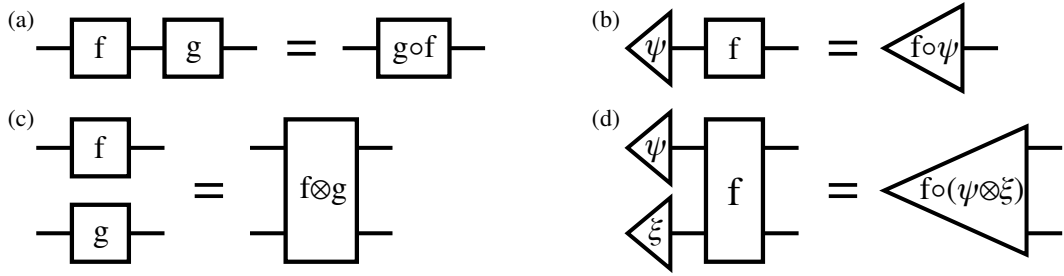}
\caption{Composition and tensor product.
(a)~Composition of operators. $(g \circ f)\indices{^a_c} = g\indices{^a_b}
f\indices{^b_c}$.
(b)~Composition of a state and an operator. $(f \circ \psi)\indices{^{a}} =
f\indices{^{a} _{b}} \psi^{b}$.
(c)~Tensor product of operators. $(f \otimes g)\indices{^{a_1 a_2}_{b_1 b_2}} = f\indices{^{a_1}_{b_1}}
g\indices{^{a_2}_{b_2}}$.
(d)~Tensor product of states composed with an operator.
$(f \circ(\psi \otimes \xi))\indices{^{a_1 a_2}} 
= f\indices{^{a_1 a_2} _{b_1 b_2}}
 \psi^{b_1} \xi^{b_2} $.}
\label{fig:comp}
\end{figure}

The category \QC{} has two composition-like operations,
the tensor product~$\otimes$,
and the composition of morphisms~$\circ$.
The composition of morphisms is represented graphically by the
horizontal juxtaposition of the corresponding
diagram elements and connecting the corresponding wires.
Likewise, tensor products of objects or morphisms are represented by
the vertical stacking of the diagram elements.
These diagrammatic structures are illustrated in
\Figref{fig:comp}.

\begin{remark}[Bifunctoriality~\cite{JS91}] In the diagrammatic calculus, the
equation 
\begin{equation}
(g \circ f) \otimes (t \circ s) = (g \otimes t) \circ (f \otimes s) 
\end{equation}
has the evident pictorial meaning which amounts to first
connecting boxes horizontally (resulting in $g \circ f$, $t \circ s$), and
then stacking them vertically to yield $(g \circ f) \otimes (t \circ s)$, or
first stacking them vertically
(resulting in $g \otimes t$, $f \otimes s$),
and then connecting the stacks horizontally to yield
$(g \otimes t) \circ (f \otimes s)$.
\end{remark}

\subsubsection{The dagger functor}

\begin{figure}
\includegraphics[width=0.65\textwidth]{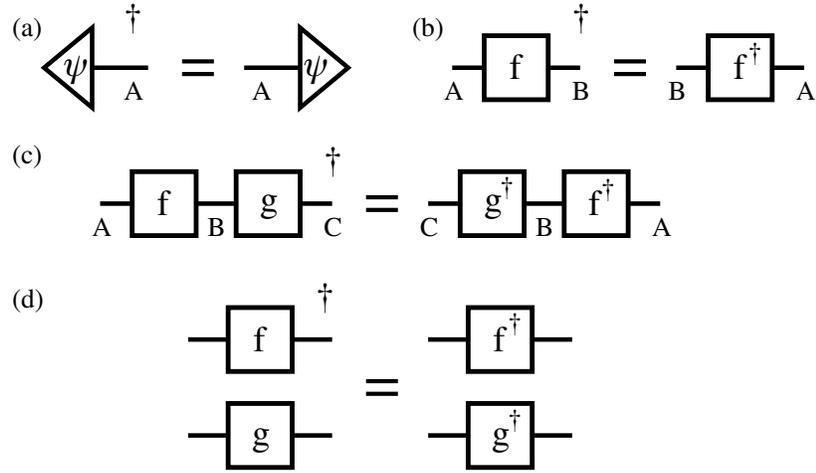}
\caption{Dagger functor.
(a)~Dagger of a state. $(\psi^\dagger)_a = \overline{\psi_a}$.
(b)~Dagger of an operator.
$(f^\dagger)\indices{^a_b} = \overline{f\indices{_b^a}}$.
(c)~Dagger of composition.
$(g \circ f)^\dagger = f^\dagger \circ g^\dagger$.
(d)~Dagger of tensor product.
$(f \otimes g)^\dagger = f^\dagger \otimes g^\dagger$.
}
\label{fig:dagger}
\end{figure}

The effect of the dagger functor on the category~\QC, taking the
Hermitian conjugate of a morphism, is represented
diagrammatically by mirroring the diagram in the horizontal direction.
Hence given a morphism~$f$, the diagrams corresponding to $f$ and
$f^\dagger$ are each others' mirror images.
The operator labels have a $\dagger$ symbol appended whereas the state and
costate symbols stay the same.
This is illustrated in \Figref{fig:dagger}.

\subsection{Cups and caps: Bell states and Bell effects}
\label{sec:cups}
We will now make use of the structure of dagger-compact
closure~\cite{baez_dolan1995} to derive
elegant dualities between morphisms of different types.
This provides an intuitive generalization of concepts surrounding the
Choi-Jamiolkowski isomorphism.

\begin{figure}[h]
\includegraphics[width=0.4\textwidth]{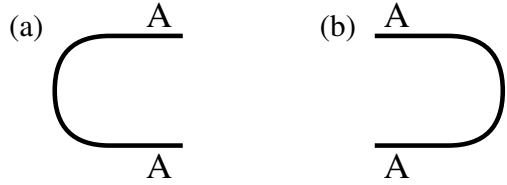}
\caption{Dagger-compact structures. (a) Cup $\eta_{A}$.  (b) Cap $\epsilon_{A}$.}
\label{fig:cup-and-cap}
\end{figure}

Building on ideas in~\cite{catQM}, we introduce two new diagrammatic
elements that do not
appear in standard quantum circuits, shown in \Figref{fig:cup-and-cap}.
In the present work, they are the only ways a wire may reverse its
direction of left-right progression.
The first one, called
a~\emph{cup}, is simply another way of denoting a state preparation
procedure for a generalized Bell state in the Hilbert
space~$\hilbert{A}^{\otimes 2}$, scaled by $\sqrt{d_A}$ where
$d_{A} = \dim \hilbert{A}$.

\begin{figure}[h]
\includegraphics[width=0.6\textwidth]{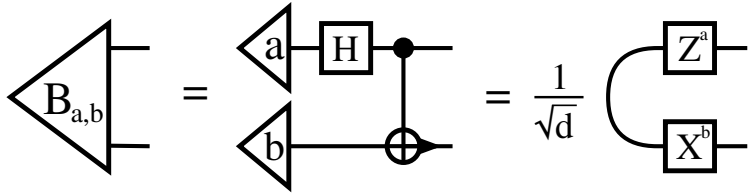}
\caption{Generalized Bell states. Preparation using the \ADD{} gate,
relation to the cup element.
}
\label{fig:bellstates}
\end{figure}

\begin{definition}[Cup]
The cup is the diagram element that corresponds to
the dagger-compact structure~$\eta$ of the category~\QC.
It is a morphism
$\eta_{A}: \eye \to A \otimes A$,
given in the computational basis as
\be
\eta_{A} := \sum_{i=0}^{d-1} \ket{i}_{A} \otimes \ket{i}_{A}
= \delta\indices{^{ij}} \ket{ij}_{A \otimes A}.
\ee
\end{definition}
It is easy to notice that $\eta_{A}$ is proportional to
the Bell state $B_{0,0}$ we defined previously,
\be
\cupket{A} :=
\ket{B_{0,0}}_{A \otimes A} =
\frac{1}{\sqrt{d_{A}}} \eta_{A},
\ee
and that the other Bell states are locally
equivalent to~$\cupket{A}$, as shown in~\Figref{fig:bellstates}.

The~\emph{cap} can be thought of physically as corresponding
to a postselected measurement (an effect) in the generalized Bell basis.
  
\begin{definition}[Cap]
The cap is the diagram element that corresponds to
the dagger-compact structure~$\epsilon$ of the category~\QC.
It is obtained by taking the dagger of the cup,
which makes it the morphism
$\epsilon_{A}: A \otimes A \to \eye$:
\be
\epsilon_{A} := \eta_{A}^\dagger
= \sum_{i=0}^{d-1} \bra{i}_{A} \otimes \bra{i}_{A}
= \delta\indices{_{ij}} \bra{ij}_{A \otimes A}.
\ee
\end{definition}

One may safely think that the purpose of these structures is to entangle
two subsystems in a way that enables very intuitive manipulation of the
corresponding circuit diagrams by bending the wires in them. This is
based on the isomorphisms they induce between states and operators.

\begin{figure}
\includegraphics[width=0.85\textwidth]{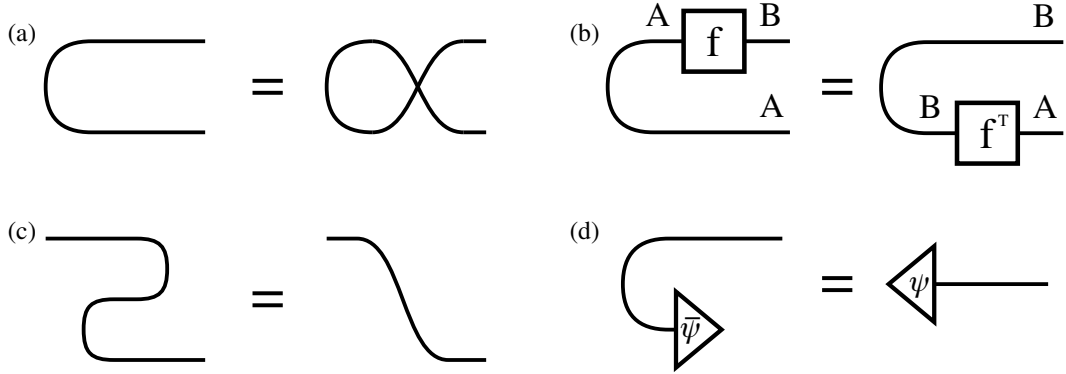}
\caption{Cup identities. (a) Symmetry.
(b) ``Sliding'' an operator around a cup transposes it in the computational basis.
(c) Snake equation.
(d) Conjugate state.
}
\label{fig:cups}
\end{figure}

Now we will demonstrate some properties of cups and caps,
corresponding to the diagram identities in~\Figref{fig:cups}.
We give proofs for cups, but corresponding identities hold for caps as
well, and the proofs can be obtained by taking the Hermitian
conjugates of the ones we give below.

\begin{theorem}[Cup and cap symmetry (\Figref{fig:cups}(a))]
Since the cup corresponds to a symmetric state, it immediately follows
that the relative order of the two subsystems is irrelevant. 
Diagrammatically this means the order of the wires can be swapped.
Cf.~\Eqref{eq:unitsymmetry}.
\end{theorem} 

\begin{theorem}[Sliding operators around cups and caps (\Figref{fig:cups}(b))]
\label{th:slide}
An operator~$f: A \to B$ can be moved (``slid'')
around a cup or a cap by transposing it in the computational basis.
Alternatively, there is an isomorphism between a cup
followed by the operator~$f$ on the first subsystem, the
state~$\ket{\opstate{f}} := \frac{1}{\sqrt{d_{A}}} \text{vec}(f^T)^k
\ket{k}_{B \otimes A}$,
and a cup followed by the operator~$f^T$ on the second subsystem.
\footnote{
The $\text{vec}$ operation takes the matrix of its operand in the
computational basis and rearranges it column by column, left to right, into a column vector.
}
\end{theorem}

\begin{proof}
\begin{align}
\notag
\left(f\indices{^i_j} \ketbra{i}{j}_1 \right) \eta_{A \, 1,2}
&=
\left(f\indices{^i_j} \ketbra{i}{j}_1 \right) \left(\delta\indices{^{kl}} \ket{k}_1 \ket{l}_2\right)
= f\indices{^i_j} \delta\indices{^{j} _k} \delta\indices{^{kl}} \ket{i}_1 \ket{l}_2\\
\notag
&= f\indices{^{ij}} \ket{i}_1 \ket{j}_2
= \text{vec}(f^T)^k \ket{k}_{1,2}
= \sqrt{d_{A}} \ket{\opstate{f}}_{1,2}
= f\indices{_j^i} \delta\indices{^{j}_l} \delta\indices{^{kl}} \ket{k}_1 \ket{i}_2\\
&
= \left(f\indices{_j^i} \ketbra{i}{j}_2 \right)
\left(\delta\indices{^{kl}} \ket{k}_1 \ket{l}_2\right)
= \left((f^T)\indices{^i_j} \ketbra{i}{j}_2 \right) \eta_{B \, 1,2}
\end{align}
\end{proof}

\begin{corollary}
All local unitary operators~$f$ are isomorphic to a state~$\ket{\opstate{f}}$ that is
locally equivalent to a generalized Bell state.
\end{corollary}

\begin{corollary}[Conversions between inputs and outputs of the same type]
More generally, a cup converts an input leg of an operator
into an output leg of the same type. The opposite is true for a cap.
\end{corollary}
\begin{proof}
 \begin{align}
\notag
\left(f \otimes \I_\omega \right) \eta_{q, \omega}
&=
\left(
\ket{b_1 \cdots b_n} \otimes \ket{x}_\omega
f\indices{^{b_1 \cdots b_n}_{a_1 \cdots a_m}}
\bra{a_1 \cdots a_m} \otimes \bra{x}_\omega
\right)
\left(\delta^{kl} \ket{k}_q \ket{l}_\omega \right)\\
\notag
&=
\ket{b_1 \cdots b_n} \otimes \ket{x}_\omega
f\indices{^{b_1 \cdots b_n}_{a_1 \cdots a_m}}
\bra{a_1 \cdots a_{q-1} a_{q+1} \cdots a_m}
\delta^{kl} 
\delta \indices{^{a_q}_k}
\delta \indices{^x_l}\\
\notag
&=
\ket{b_1 \cdots b_n} \otimes \ket{a_q}_\omega
f\indices{^{b_1 \cdots b_n}_{a_1 \cdots a_{q-1}}^{a_q}_{a_{q+1} \cdots
a_m}}
\bra{a_1 \cdots a_{q-1} a_{q+1} \cdots a_m}\\
&=:
\ket{b_1 \cdots b_n} \otimes \ket{a_q}_\omega
\hat{f}\indices{^{b_1 \cdots b_n a_q}_{a_1 \cdots a_{q-1} a_{q+1} \cdots a_m}}
\bra{a_1 \cdots a_{q-1} a_{q+1} \cdots a_m}.
\end{align}
\end{proof}

\begin{theorem}[Snake equation (\Figref{fig:cups}c)]
A cup and a cap can combine to cancel each other.  In other words
a double bend in a wire can be pulled straight.
In Section~\ref{sec:examples} we show how this operation actually
corresponds to the standard quantum teleportation
protocol~\cite{catQM}.
\end{theorem}
\begin{proof}
\begin{align}
\notag
\left(\epsilon_{1,2} \otimes \I_3 \right)
 \left(\I_1 \otimes \eta_{2,3} \right)
&= \left(\delta\indices{_{ij}} \bra{i}_1 \bra{j}_2 
\otimes \I_3 \right)
  \left(\I_1 \otimes \delta\indices{^{kl}} \ket{k}_2
\ket{l}_3\right)\\
&= \delta\indices{_{ij}} \delta\indices{^{kl}}
\delta\indices{^j _k} \ket{l}_3 \bra{i}_1 
= \ket{i}_3 \bra{i}_1 = \I_{3,1}.
\end{align}
\end{proof}

\begin{theorem}[Conjugate states (\Figref{fig:cups}d)]
Cups and caps induce an isomorphism between states
$\ket{\psi}~=~\psi^k~\ket{k}$ and their
\emph{conjugate states} $\bra{\bar{\psi}}~:=~\bra{k}~\psi_k$,
which are obtained by complex conjugating
the coefficients of the corresponding bra in the computational basis.
\end{theorem}
\begin{proof}
 \begin{align}
\bra{\bar{\psi}}_2 \; \eta_{1,2}
= (\psi_j \bra{j}_2) \left(\delta\indices{^{kl}}
\ket{k}_1 \ket{l}_2\right)
= \psi_j \delta\indices{^j _l} \delta\indices{^{kl}}
\ket{k}_1
= \psi^k \ket{k}_1 = \ket{\psi}_1.
\end{align}
\end{proof}

\begin{remark}[Diagrammatic adjoints]
As mentioned above, cups and caps allow us to take the transpose~$f^T$ of a
linear map~$f$. Following the (now) standard string diagram literature we
introduce the derived concept of adjoint $\overline{f}$ (see \eg~\cite[Sec. XIV.2]{kassel}):
\begin{center}
  \includegraphics[width=0.7\textwidth]{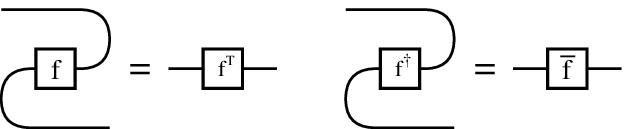}
\end{center}
\end{remark}

\begin{remark}[Basis dependence of transposition and complex conjugation]
At first it might seem strange that we should encounter basis-dependent
operations such as transposition and complex conjugation. However,
this is a direct result of us having chosen a preferred computational basis
and defined the cup/cap operators in terms of it.
\end{remark}

\subsection{Symmetric dots}

In this subsection we consider a special class of morphisms
we call~\emph{symmetric dots}, or just dots for short.\footnote{
Note that other authors use the term dot to mean different kinds of morphisms
with certain well-defined properties. In the present paper, however, it
always refers to a symmetric dot.}
Most importantly, the definition for each kind of dot is readily
extensible to an arbitrary number of input and output legs, all of
which have the same, arbitrary dimension.

Symmetric dots are defined by their \emph{kind}, \emph{order}, \emph{dimension},
and \emph{color}.
An dot corresponding to a morphism of the type
$A^{\otimes m} \to A^{\otimes n}$
is of the order~$(m,n)$ and dimension~$d = \dim A$.
The kind of a dot defines its effect, and color the basis in which it
operates.

In the diagrams, dots are represented by a circular node (``dot'') with a
symbol denoting the kind of the dot.
By default the dots operate in the computational basis.
If this is not the case, the basis is specified by
a label next to the dot symbol.
Mathematically, a dot of the kind~$D$ and order~$(m,n)$, operating in the basis~$B$,
is denoted as~\DOT{B}{m \to n}.
Again, if $B$~is omitted, the dot is assumed to operate in the
computational basis.

\begin{figure}[h]
\includegraphics[width=0.5\textwidth]{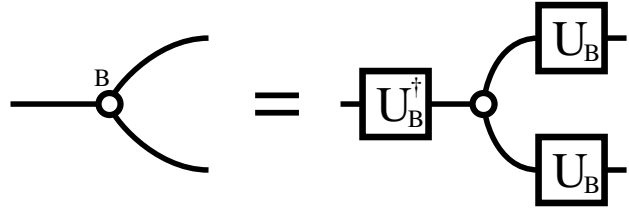}
\caption{Changing the color of a dot from the computational basis to the
orthonormal basis $B = \{ \ket{b_k} \}_k$
using the unitary operator $U_B = \sum_k \ketbra{b_k}{k}$.} 
\label{fig:dotcolor}
\end{figure}

Color change occurs when a dot is rotated from one basis into another.
For example, the unitary transformation
$U_B := \sum_k \ketbra{b_k}{k}$
from the computational basis to the
orthonormal basis $B = \{ \ket{b_k} \}_k$
can be used to convert any dot \DOT{}{}
into~\DOT{B}{}:
\be
\DOT{B}{m \to n} = U_B^{\otimes n} \DOT{}{m \to n} U_B^{\dagger \otimes m}.
\ee
This is illustrated in~\Figref{fig:dotcolor}.

\begin{figure}[h]
\includegraphics[width=1\textwidth]{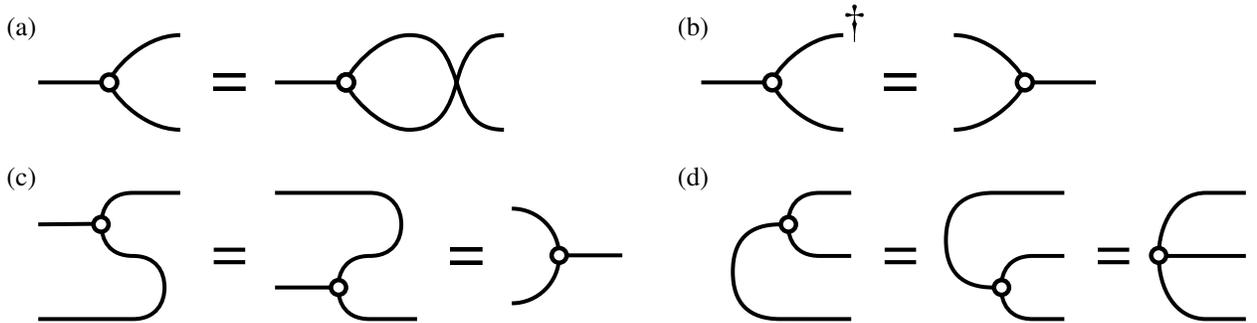}
\caption{Symmetric dots. We use a generic $1$-to-$2$ dot as an example, but
these properties apply to dots of all kinds, orders, dimensions and colors.
(a) Permutation symmetry of the input and output legs (S1).
(b) Dagger reverses the order but preserves the other properties of the dot (S2).
(c,d) Cups and caps can be used to bend inputs into outputs and vice versa (S3).
}
\label{fig:dots}
\end{figure}

Furthermore,
symmetric dots are required to have the following properties:
\begin{itemize}
\item[S1]
They are invariant under all permutations of their input and output legs.

\item[S2]
The dagger functor simply converts a dot's input legs
into output legs and vice versa, preserving all the other properties:
\be
\DOT{B}{m \to n \dagger} = \DOT{B}{n \to m}.
\ee

\item[S3]
Their legs can be ``bent'' using cups and caps.
Attaching a cup(cap) to an input(output) leg converts it into an
output(input) leg, respectively:
\begin{align}
\notag
\DOT{B}{m \to n} \: \eta &= \DOT{B}{(m-1) \to (n+1)},\\
\epsilon \: \DOT{B}{m \to n} &= \DOT{B}{(m+1) \to (n-1)}.
\end{align}
Unlike S1 and S2, this property places a restriction on the colors the dot can
appear in; we must have~$U_B = \overline{U_B}$. 
\end{itemize}
These symmetry properties are illustrated in~\Figref{fig:dots}.

\begin{definition}[Pruning element]
A \emph{pruning element}~$\ket{\prune_D}$ for a dot~\DOT{}{} is a state/costate which, when
connected to a leg of the dot eliminates that leg, reducing the
corresponding order of the dot by one:
\begin{align}
\notag
\DOT{}{m \to n} \ket{\prune_D} &= \DOT{}{(m-1) \to n},\\
\bra{\prune_D} \DOT{}{m \to n} &= \DOT{}{m \to (n-1)}.
\end{align}
\end{definition}

Now we will introduce specific kinds of symmetric dots.

\subsubsection{Copy dots}

\begin{figure}[h]
\includegraphics[width=0.3\textwidth]{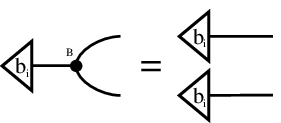}
\caption{Copy dot \COPY{B}{1 \to 2} in the orthonormal basis $B = \{ \ket{b_k} \}_k$.
}
\label{fig:copy}
\end{figure}

\begin{definition}[\COPY{}{}]
The $m$-to-$n$ copy dot is defined in the computational basis as
\be
\label{eq:copymerge}
\COPY{}{m \to n} := \sum_k \ketbra{\underbrace{k \cdots k}_{n}}{\underbrace{k \cdots k}_{m}}.
\ee
In our diagrammatic notation a copy dot is represented by a
simple black dot~$\bullet$.
Connecting a basis state $\ket{k}$ (or the corresponding costate) to
any of the legs of the copy dot collapses the sum and breaks
the dot up into unconnected copies of~$\ket{k}$ and~$\bra{k}$.
For example the $1$-to-$2$ copy dot
\be
\COPY{B}{1 \to 2} = \sum_k \ketbra{b_k b_k}{b_k}, 
\ee
given the state $\ket{b_k}$ as the input,
produces two copies of the same state as output.
\footnote{This does not violate the no cloning theorem since the operator
can only faithfully copy a single fixed basis.}
\Figref{fig:copy} depicts this in diagram form.
\end{definition}

\begin{remark}[\COPY{}{} as a quantum operation]
Direct calculation gives
\be
\COPY{B}{m \to n \dagger} \COPY{B}{m \to n} 
= \sum_{i} \ketbra{\underbrace{b_i \cdots b_i}_{m}}{\underbrace{b_i \cdots b_i}_{m}}
= \I \quad \Leftrightarrow \quad m=1.
\ee
Hence an $m$-to-$n$ copy dot is a valid quantum operation iff $m=1$.
This property is not preserved under the dagger ---
$\COPY{B}{1 \to n \dagger} = \COPY{B}{n \to 1}$ (\emph{merge})
is not a valid quantum operation if $n > 1$.
It is however still useful to
consider its properties --- by invoking arguments such as 
postselection it can be given a physical meaning.  
\end{remark}

\subsubsection{Plus dots}
\begin{definition}[\PLUS{}{}]
We define the $m$-to-$n$ plus dot in the computational basis as
\begin{align}
\label{eq:plusdot}
\notag
\PLUS{}{m \to n} :=& H^{\otimes n} \COPY{}{m \to n} H^{\otimes m}
= \COPY{x}{m \to n} \NEG^{\otimes m}\\
=& \frac{1}{d^{(m+n-2)/2}}
\sum_{\substack{r_1 \cdots r_m \\ s_1 \cdots s_n}}
\delta_{\left(\sum_i r_i \oplus \sum_j s_j\right), 0}
\ket{s_1 \cdots s_n} \bra{r_1 \cdots r_m},
\end{align}
where $d$~is the dimension of the legs.
Roughly speaking, the plus dot ensures all its inputs and outputs in
the given basis sum to zero $\mod d$.
Note that the plus dot is not a copy dot unless~$d = 2$.
Diagrammatically, a plus dot is represented by a circular node with a
plus symbol inside.\footnote{
The diagrammatic representation of \PLUS{}{1 \to 1} (a wire with
$\oplus$ on it) must not be confused with the notation occasionally
used in standard QCDs, in which $\oplus$ denotes the \NOT{} gate.
We instead have $\PLUS{}{1 \to 1} = \NEG$.
Following this logic we could have used~$\oplus$ as the symbol of the
\NEG{} gate as well but felt this notational parsimony would not have
been worth the potential for confusion.
}
\end{definition}

\subsubsection{Simplification rules for \COPY{}{} and \PLUS{}{}}

Neighboring copy dots of the same color can be merged into a single dot.
In categorical quantum mechanics, this is called the ``spider law''~\cite{redgreen}.
We shall provide a slightly more general version below.
\begin{definition}[Spider law]
Dots of the kind \DOT{}{} are said to fulfill the spider law
with the operator~$G$ as the \emph{glue}
iff
any connected graph with $m$~inputs and $n$~outputs comprised
of \DOT{}{} dots of the same dimension, with $G$s on
all the internal legs connecting two neighboring dots,
can be equivalently expressed as a single~\DOT{}{m \to n}~dot,
as shown in~\Figref{fig:spider}.
\begin{figure}[h]
\includegraphics[width=0.7\textwidth]{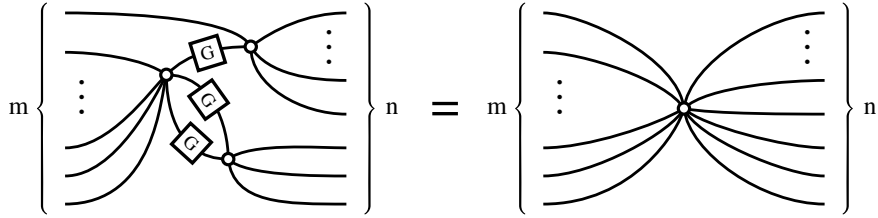}
\caption{Spider law for a generic dot with the operator~$G$ working as the glue.}
\label{fig:spider}
\end{figure}

The spider law is colorblind;
if \DOT{}{} fulfills the spider law with~$G$ as the glue,
then \DOT{B}{} will fulfill the spider law with
$G_B := U_B \: G \: U_B^\dagger$ as the glue.
Furthermore, the symmetry property S2 requires that if
$G$~works as glue for \DOT{}{} then so does~$G^\dagger$ .
\\
\\
Corollary:
For any dot~$D$ obeying the spider law with $G$~as the glue,
$G D^{0 \to 1}$ functions as a
pruning element.
\end{definition}

\begin{theorem}[Spider law] 
Copy dots obey the spider law with the trivial glue (identity).
Plus dots obey the spider law with the negation gate~$\NEG$ as the glue.
\end{theorem}

\begin{figure}[p]
\includegraphics[width=0.95\textwidth]{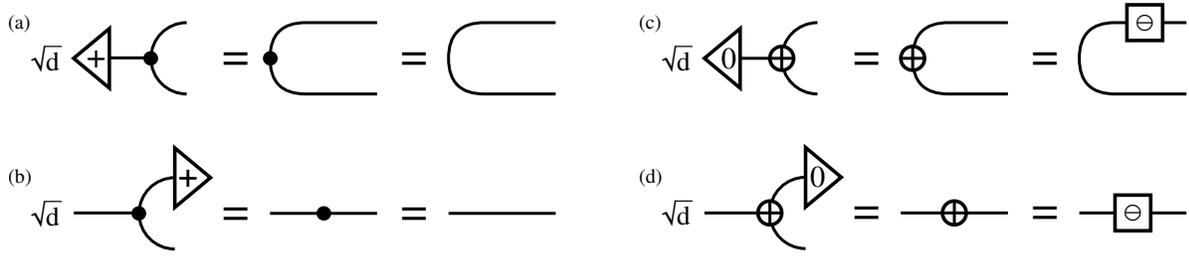}
\caption{Pruning elements for the (a,b) $\bullet$ and (c,d) $\oplus$ dots.
(In a qubit system the~$\NEG$ gate reduces to identity.)
} 
\label{fig:dotprune}
\end{figure}

\begin{theorem}[Pruning elements]
$\sqrt{d} \; \ket{0}$ is a pruning element for \PLUS{}{}
as it makes the
corresponding index vanish in the Kronecker delta
in \Eqref{eq:plusdot}.
Using this result, it follows that $\sqrt{d} \; \ket{+}$ is a pruning
element for \COPY{}{}.
(See~\Figref{fig:dotprune}.)
\end{theorem}

\begin{figure}[p]
\includegraphics[width=0.75\textwidth]{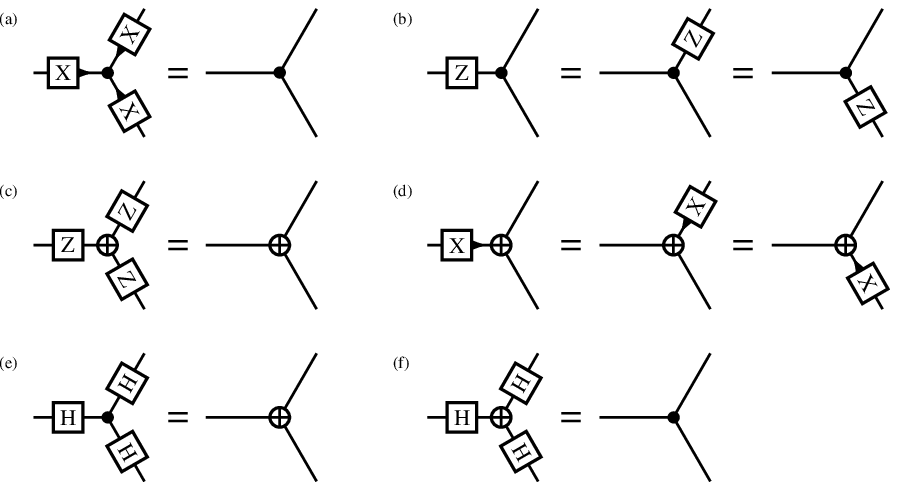}
\caption{Commutation rules for (a,b) the $\bullet$ dot and
(c,d) the $\oplus$ dot with the $Z$ and $X$ gates.  (e,f) Conversions between $\bullet$ and $\oplus$ dots using discrete
Fourier transform gates~$H$. Analogous rules apply to $\bullet$ and $\oplus$ dots with an arbitrary
number of legs. \footnote{ Aesthetic interlude:
Given the computational basis, the set of operations
$\{1, ^T, \overline{\phantom{g}}, ^\dagger\}$ is isomorphic to the Klein group~$Z_2 \times Z_2$,
which also is the symmetry group of the rectangle.
We can use this to illustrate the symmetry
properties of operators with their symbols, by equating~$^\dagger$
with horizontal reflection  of the symbol,
$^T$~with a 180-degree rotation (\eg{} sliding the symbol around a cup/cap!),
and $\overline{\phantom{g}}$ with vertical reflection.
By adding an arrow to the $X$ gate symbol to denote the
direction of incrementation
the symmetry properties of the $Z$ and $X$ gates are
represented by the symmetries of their gate symbols:
$Z^T = Z$, $\overline{X} = X$.
(This is analogous to the function of the corner marker on
the morphism symbols in~\cite{Selinger09}.)
The arrow would not be necessary if the $X$ gate symbol
had the correct symmetry in itself (like the letter $E$, for
example), but we chose to go with the more traditional choice.}}
\label{fig:copyZX}
\end{figure}

\begin{figure}[p]
\includegraphics[width=0.6\textwidth]{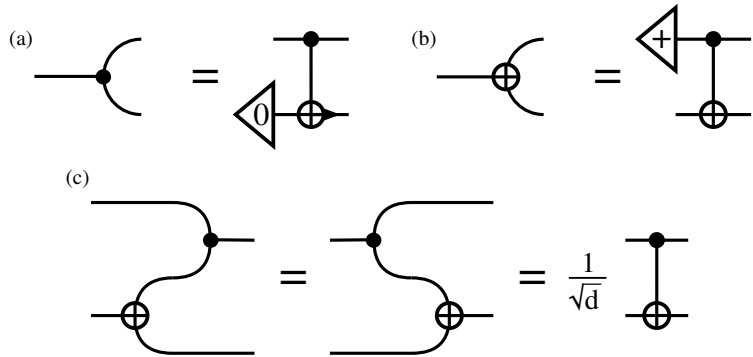}
\caption{Connection between the copy and plus dots and the \NADD{} gate.
(a,b) Explicit constructions for the 
$\bullet$ and $\oplus$ dots. (c) Combining the $\bullet$ and $\oplus$ dots
yields \NADD.}
\label{fig:dots_nadd}
\end{figure}

\begin{theorem}[Commutation rules for the $Z$ and $X$ gates with dots]
Since the $Z$~gate shares its eigenbasis with \COPY{}{}, they
fully commute:
\be
\sum_k \ket{k}_1 \ket{k}_2 (\bra{k}_1 Z)
= \sum_k (Z \ket{k}_1) \ket{k}_2 \bra{k}_1
= \sum_k \ket{k}_1 (Z \ket{k}_2) \bra{k}_1.
\ee
The $X$~gate, however, is multiplied when it passes a \COPY{}{}:
\be
\sum_k \ket{k}_1 \ket{k}_2 (\bra{k}_1 X)
= \sum_k \ket{k}_1 \ket{k}_2 \bra{k \ominus 1}_1
= \sum_k \ket{k \oplus 1}_1 \ket{k \oplus 1}_2 \bra{k}_1
= \sum_k (X \ket{k}_1) (X \ket{k}_2) \bra{k}_1
\ee
One obtains equivalent results for \PLUS{}{}, with the roles of $Z$ and $X$
exchanged. These commutation rules are presented in
\Figref{fig:copyZX}.
Even though we used $1 \to 2$ dots in our proofs above, 
analogous rules apply to \COPY{}{} and \PLUS{}{} dots with an arbitrary
number of legs.
\end{theorem}

Now we have assembled all the necessary ingredients to make the dots
do something useful.
As the astute reader probably already has noticed,
the notation we use for the $\bullet$ and $\oplus$ dots is
suggestively similar to the \NADD{} gate symbol, for a good reason.
\Figref{fig:dots_nadd} shows how the \NADD{} gate can be built out of
dots, and how the $\bullet$ and $\oplus$ dots can in some cases be
explicitly constructed using the \NADD{} gate.

\begin{figure}[h]
\includegraphics[width=0.9\textwidth]{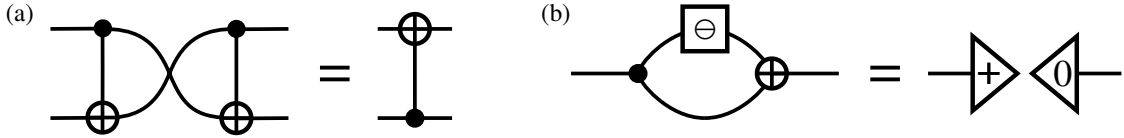}
\caption{(a) Bialgebra law. (b) Hopf law.}
\label{fig:bihopf}
\end{figure}

\begin{theorem}[Bialgebra law]
Two \NADD{} gates connected to each other via a \SWAP{} gate
as in~\Figref{fig:bihopf}(a) are equal to a single inverted \NADD{} gate.
A~similar equality holds for the $\bullet$ and $\oplus$ dots with an
extra factor of~$\sqrt{d}$ on the left.
\begin{proof}
\begin{align}
\notag
\NADD_{1,2} \SWAP_{1,2} \NADD_{1,2}
&= \sum_{abxy} \ket{x,\ominus x \ominus y}_{1,2} \braket{y,x}{a,\ominus a \ominus b} \bra{a,b}_{1,2}\\
&= \sum_{ab} \ketbra{\ominus a \ominus b, b}{a,b}_{1,2}
= \NADD_{2,1}.
\end{align}
\end{proof}
\end{theorem}

\begin{theorem}[Hopf law]
The \COPY{}{} and \PLUS{}{} dots fulfill the Hopf law~\cite{hopf} with the~\NEG{}
gate as the antipode, as shown in~\Figref{fig:bihopf}(b).
\begin{proof}
\begin{align}
\notag
\PLUS{}{2 \to 1} \NEG_{1} \COPY{}{1 \to 2}
&= \frac{1}{\sqrt{d}}
\sum_{kabcx} \delta_{a \oplus b \oplus c, 0}
\ket{a} \bra{bc} (\ketbra{x}{\ominus x} \otimes \I) \ket{kk} \bra{k}\\
&= \frac{1}{\sqrt{d}}
\sum_{ka} \delta_{a \ominus k \oplus k, 0}
\ket{a} \bra{k}
=
\ket{0}
\left( \frac{1}{\sqrt{d}} \sum_k \bra{k} \right)
= \ket{0} \bra{+}.
\end{align}
\end{proof}
\end{theorem}

\subsection{From diagrams to quantum operations}

The extended QCDs each correspond to a \QC-morphism.
However, not every such morphism is physically implementable on its own.
In quantum mechanics a state operator can (in principle) undergo any
evolution that can be expressed as a linear, completely positive map (CPM).
The mapping from \QC-morphisms to CPMs is easiest achieved using the
operator-sum representation, in which each morphism corresponds to a
Kraus operator.

\begin{definition}[Complete set of \QC-morphisms]
We call a set of \QC-morphisms $S = \{f_i\}_i \subset \Hom[\QC]{A}{B}$
\emph{complete} iff it
corresponds to a quantum operation, that is,
\be
S \:\: \text{is complete} \quad \Leftrightarrow \quad \sum_i f_i^\dagger f_i = 1_A.
\ee
The effect of $S$ on the state operator $\rho: A \to A$
is
\be
\rho \mapsto \sum_i f_i \rho f_i^\dagger.
\ee
\end{definition}
Another category-based approach to representing CPMs using diagrams can be found in~\cite{Selinger07}.

\begin{theorem}[Properties of complete sets of \QC-morphisms]
The following properties immediately follow from the definition:
\begin{enumerate}
\item[(a)]
If $f: A \to A$ is unitary, it is complete on its own.

\item[(b)]
A state $\psi: \eye \to A$ is complete on its own
iff it is normalized: $\braket{\psi}{\psi} = 1$.

\item[(c)]
A set of costates $\{\chi_k: A \to \eye\}_k$ is complete
if the corresponding states form an orthonormal basis for~$A$: $\sum_k
\ketbra{\chi_k}{\chi_k}_{A} = 1_{A}$. In this case the set of costates
corresponds to a projective measurement in this basis.

\item[(d)]
If $\{f_i\}_i \subset \Hom[\QC]{A}{B}$ and
$\{g_j\}_j \subset \Hom[\QC]{C}{D}$ are complete sets of morphisms,
the tensor product set
$\{f_i \otimes g_j\}_{ij} \subset \Hom[\QC]{A \otimes C}{B \otimes D}$ is also complete.

\item[(e)]
If $\{f_i\}_i \subset \Hom[\QC]{A}{B}$ and
$\{g_j\}_j\subset \Hom[\QC]{B}{C}$ are complete sets of morphisms,
the composed set $\{g_j~\circ~f_i\}_{ij} \subset \Hom[\QC]{A}{C}$ is also complete.
\end{enumerate}
\end{theorem}

\begin{figure}[h]
\includegraphics[width=0.9\textwidth]{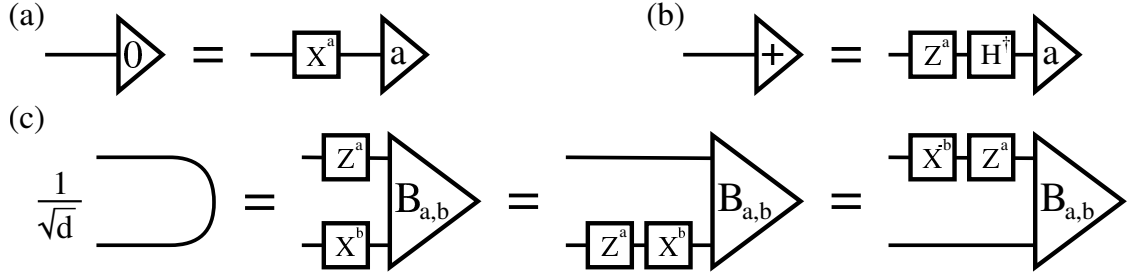}
\caption{Representations of the (a) $\bra{0}$ and (b) $\bra{+}$ costates and (c) caps in
  terms of complete sets of costates with local corrections.
  The generalized Bell costates~$\bra{B_{a,b}}$
  can be presented in terms of computational basis costates
  \eg{} using the inverse of the circuit in \Figref{fig:bellstates}.}
\label{fig:impl}
\end{figure}

In constructing complete sets of \QC-morphisms it is useful to be able to
implement caps and certain other costates in terms of 
projective measurements followed by local unitary corrections
dependent on the outcome.
This can be done by first expressing the costate in terms of
a complete set of standard basis costates
(as shown in \Figref{fig:impl}) and then 
using Theorem~\ref{th:slide} together with commutation rules between
various circuit elements and the $Z$ and~$X$ gates.
In the computational basis $Z^T = Z$ and $X^T = X^{-1}$, so they
both can readily be slid around cups and caps. By shuttling them
along the circuit to positions which causally follow the
costate that introduced them (if possible!), the circuit becomes physically
implementable.
Examples on how this is accomplished in practice are given in the next section.

\section{Applications}

Here we present some applications of the extended quantum
circuit diagram methods derived in the last section.
First we give some examples of circuit simplification, and then
derive several well-known quantum protocols for systems of arbitrary
dimension using almost no algebra beyond what is implicit in the diagrams.

\subsection{Circuit simplification}

\begin{example}[Commuting $Z$ and $X$ gates through a \NADD{} gate]
Start by breaking the \NADD{} gate into a copy dot and a plus dot as
shown in \Figref{fig:dots_nadd}.
Then apply the commutation rules presented in \Figref{fig:copyZX}
to commute the $Z$ and $X$ gates through the dots one by one, and finally put the
\NADD{} gate together again.
The result is the $d$-dimensional generalization of the familiar
commutation rules between $\sigma_z$, $\sigma_x$ and a \CNOT.
\begin{center}
\includegraphics[width=1\textwidth]{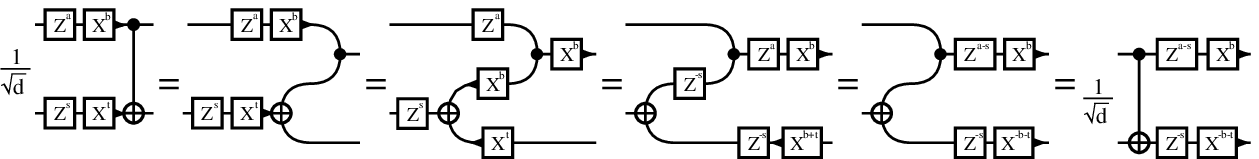}
\end{center}
\label{ex:ZX_NADD}
\end{example}

\begin{example}[GHZ circuit]
We are given the $d$-dimensional version of the standard circuit for
preparing Greenberger-Horne-Zeilinger (GHZ) states.
We start by breaking up the \ADD{} gates into $\bullet$ and $\oplus$ dots
as shown in \Figref{fig:dots_nadd}(c),
then apply the pruning element identities in~\Figref{fig:dotprune}
(or alternatively use the dot constructions in
\Figref{fig:dots_nadd}(a,b) in reverse)
to obtain a network of copy dots and cups. We use the cups to bend the
input legs of the copy dots into output legs, and finally invoke the
spider law to fuse the copy dots together:
\begin{center}
\includegraphics[width=0.7\textwidth]{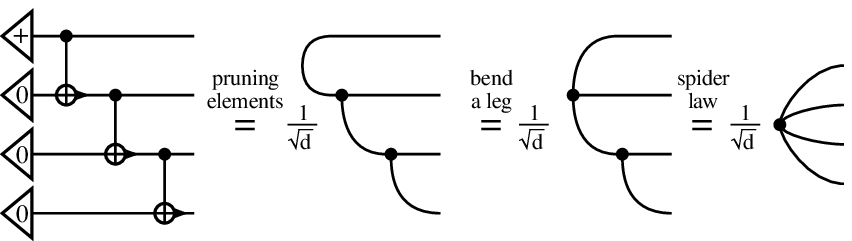}
\end{center}
The result is a scaled \COPY{}{0 \to 4} dot, which is equal to
the four-qudit GHZ state as expected:
\be
\frac{1}{\sqrt{d}} \COPY{}{0 \to 4} = \frac{1}{\sqrt{d}} \sum_k \ket{kkkk} = \ket{\text{GHZ}_4}.
\ee
\label{ex:ghz}
\end{example}

\subsection{Quantum protocols}
\label{sec:examples}

\begin{example}[Superdense coding]
We start with a diagram representing a cup state followed by local
operation $Z^p X^{-q}$ by Alice, and
finally a Bell measurement with the outcome $(a,b)$ by Bob.
We then express the Bell costate using a cap and $Z$ and $X$ gates,
slide the gates around the bends and obtain a trace expression which
is easily evaluated using~\Eqref{eq:ZXtrace}.
\begin{center}
\includegraphics[width=0.9\textwidth]{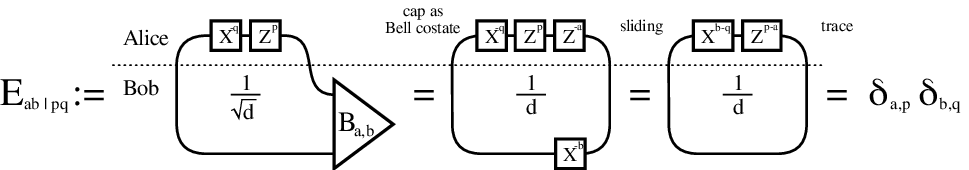}
\end{center}
The corresponding Kraus operators are
$E_{ab|pq} = \delta_{a,p} \delta_{b,q}$.
This set of morphisms is complete for all possible local
operations~$(p,q)$.
Furthermore, the probability of Bob obtaining the measurement outcome
$(a,b)$ is
$P_{ab} = \Tr(E_{ab|pq} \rho E_{ab|pq}^\dagger) = \delta_{a,p}
\delta_{b,q} \Tr(\rho) = \delta_{a,p} \delta_{b,q}$.
Hence the result of Bob's measurement is completely determined by
Alice, and she can use this protocol to transmit two d-its worth of information to Bob.
\end{example}

\begin{example}[Teleportation~\cite{PhysRevLett.70.1895}]
Starting with a scaled identity morphism, we first
use the snake equation, then express the cap in terms of a Bell costate
preceded by $Z$ and $X$ gates, and slide the gates around the cup.
This gives us a causal diagram that represents the $(a,b)$ outcome of a Bell
measurement by Alice, followed by local corrections by Bob dependent on the
measurement result.
\begin{center}
\includegraphics[width=0.9\textwidth]{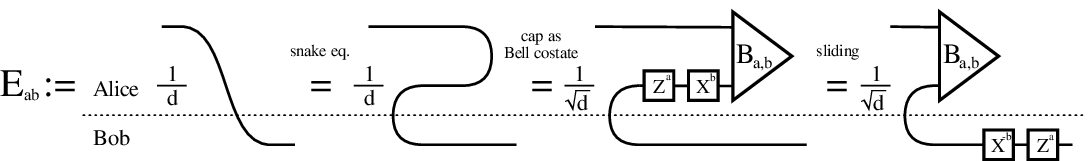}
\end{center}
The corresponding Kraus operators are
$E_{ab} = \frac{1}{d} \I$ for all $(a,b)$.
This set of morphisms is easily seen to be complete. Hence
together these diagrams must represent a physical operation,
$\rho \mapsto \sum_{ab} E_{ab} \rho E_{ab}^\dagger = \rho$,
which faithfully transports any quantum state $\rho$ from Alice to Bob.
\end{example}

\begin{example}[Teleportation through a gate~\cite{tp_through_gate}]
Starting from a two-qudit gate (\NADD{} in our example),
we use the teleportation protocol once for each input qudit,
commute the local $Z$ and $X$ corrections through the \NADD{} as
in Example~\ref{ex:ZX_NADD}, and finally regroup the gates.
\begin{center}
\includegraphics[width=0.9\textwidth]{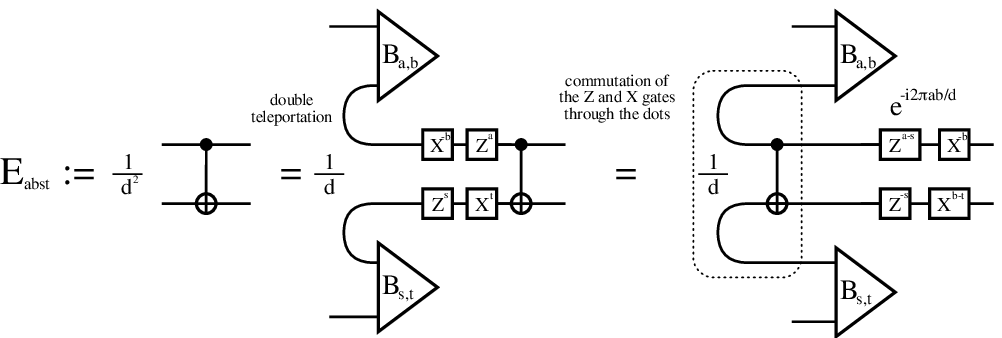}
\end{center}
The resulting diagrams each correspond to the same \NADD{} gate
operation and form a complete set.
This allows us to implement any gate~$U$ in an atemporal order:
First we apply the gate to a number of cup states, obtaining 
the state~$\ket{\opstate{U}}$
(inside the dotted line in the diagram), isomorphic to~$U$.
The inputs are then teleported ``through'' the gate-state, effectively
applying $U$ on them, even if they did not even exist yet when the
gate was actually used.
Furthermore, the states~$\ket{\opstate{U}}$ can be prepared beforehand
in large numbers and used only when
needed. This is useful \eg{} in the case where the success of an
individual $U$~operation is not guaranteed, but the computation itself
must not fail.
\end{example}


\section{Conclusion}
\label{sec:conclusion}

We have constructed a categorical framework which extends the quantum circuit model
by providing an explicit realization of a dagger-compact category
that can model finite-dimensional quantum systems,
and introduced its key algebraic properties diagrammatically.
Our construction explicitly allows for the interaction of systems of
arbitrary, possibly unequal, dimensions and thus can handle generic
tensor network states, including matrix product states.
We anticipate that our approach has further applications in applying
category theory and related ideas to tensor networks, as was
initiated in recent work~\cite{biamonte2010}.

\begin{acknowledgments}
We thank Samson Abramsky and John Baez. JDB received
support from EPSRC grant EP/G003017/1.
VB visited Oxford using funding from this same grant.
JDB completed large parts of
this work visiting the Center for Quantum Technologies, at
the National University of Singapore (hosted by Vlatko
Vedral).  
\end{acknowledgments}

\appendix

\section{Categories}
In this section we will sketch the basic concepts and definitions that
surround the present work
for the benefit of readers unfamiliar with category theory,
in a way that we hope appeals to researchers working on quantum
information.
We will skim over details not essential to the present study.
For a more complete treatment of the subject, see \eg{}~\cite{Selinger09}.

\begin{definition}[Category]
A \emph{category}~\cat{C} is an algebraic structure that consists of
\begin{enumerate}
\item[(1)]
$\Ob{C} = \{A, B, C, \ldots\}$, a class of \emph{objects}.

\item[(2)]
$\text{hom}(\cat{C})$, a class of
\emph{morphisms} (sometimes called arrows), that is,
maps between the objects.\\
For every pair of objects $A,B \in \Ob{C}$ we use
$\Hom{A}{B} \subset \text{hom}(\cat{C})$
to denote the set of morphisms from~$A$ to~$B$ in the category.

\item[(3)]
\emph{compositions} of morphisms, \ie{},
for every triple of objects $A,B,C$, the binary operation
\[
\circ:~\Hom{B}{C}~\times~\Hom{A}{B}~\to~\Hom{A}{C}.
\]
\end{enumerate}
Furthermore, the components of \cat{C} must fulfill the following axioms:
\begin{enumerate}
\item[(i)]
Associativity of composition: $(h \circ g) \circ f = h \circ (g \circ f)$
holds for all morphisms
$f \in \Hom{A}{B}$, $g \in \Hom{B}{C}$, $h \in \Hom{C}{D}$.

\item[(ii)]
Existence of identity morphisms:
For every object $A \in \Ob{C}$ there is an identity
morphism $1_A~\in~\Hom{A}{A}$ such that for every morphism
$f \in \Hom{A}{B}$ we have
$1_B \circ f = f \circ 1_A = f$.
(It can readily be shown that the identity morphisms are unique.)
\end{enumerate}

An \emph{isomorphism} is an invertible morphism.
The map $f \in \Hom{A}{B}$ is an isomorphism iff $\exists g \in \Hom{B}{A}$
for which
$g \circ f = 1_A$ and $f \circ g = 1_B$. This makes $f$ and $g$ each
others' inverses: $g = f^{-1}$.
\end{definition}

\begin{definition}[Functor]
Given categories $\cat{C}$ and~$\cat{D}$, a \emph{covariant functor}
$F: \cat{C} \to \cat{D}$ is a structure-preserving mapping between them.
More specifically, it consists of
\begin{enumerate}
\item[(1)]
a function
$F: \Ob{C} \to \Ob{D}$

\item[(2a)]
for every pair of
objects $A, B \in \Ob{C}$, a function
$\F{AB}: \Hom{A}{B} \to \Hom[D]{F(A)}{F(B)}$.\footnote{
We may use the same name~$F$ for all the functions involved in the definition
of a functor~$F$ since the
appropriate one can always be deduced from the context.
}
These functions must preserve the compositional structure of the
category; we must have
\[
\F{AC}(g \circ f) = \F{BC}(g) \circ \F{AB}(f) \quad \forall f \in
\Hom{A}{B}, \forall g \in \Hom{B}{C}.
\]
This property, together with the uniqueness of the identity morphisms gives
$\F{AA}(1_A) = 1_{F(A)}$.
\end{enumerate}
A \emph{contravariant} functor is like a covariant one, except it
reverses the directions of the morphisms:
\begin{enumerate}
\item[(2b)]
for every pair of objects $A, B \in \Ob{C}$, a function
$\F{AB}: \Hom{A}{B} \to \Hom[D]{F(B)}{F(A)}$.
In this case we must have
\[
\F{AC}(g \circ f) = \F{AB}(f) \circ \F{BC}(g) \quad \forall f \in
\Hom{A}{B}, \forall g \in \Hom{B}{C}.
\]
\end{enumerate}
An \emph{endofunctor} is a functor from a category to itself.
\end{definition}

We will now build on these basic definitions in several key stages.
The first is the notion of a \emph{monoidal} category, which is a category
equipped with a \emph{tensor product}~$\otimes$.

\begin{definition}[Monoidal category]
A \emph{monoidal} category~\cat{C} is a category equipped with
\begin{enumerate}
\item[(1)]
a covariant bifunctor called the \emph{tensor product}, $\otimes: \cat{C} \times \cat{C} \to \cat{C}$,
which typically uses the infix notation,

\item[(2)]
a \emph{unit object}~$\eye \in \Ob{C}$, and

\item[(3)]
three families of natural isomorphisms:
the \emph{associators} $\alpha$
and the \emph{left and right unitors} $\lambda$ and $\rho$.

The associators define the associativity of the tensor product.
Their naturality requires the following diagram to commute for all
$f,g,h \in \text{hom}(\cat{C})$:
\[
\begin{CD}
(A \otimes B) \otimes C @>\alpha_{A,B,C}>> A \otimes (B \otimes C)\\
@V(f \otimes g) \otimes hVV  @VVf \otimes (g \otimes h)V\\
(A' \otimes B') \otimes C' @>>\alpha_{A',B',C'}> A' \otimes (B' \otimes C')
\end{CD}
\]

The left and right unitors define the behavior of the tensor product
with respect to the unit object; for all $f \in \text{hom}(\cat{C})$,
the following diagrams must commute:
\[
\begin{CD}
\eye \otimes A @>\lambda_{A}>> A\\
@V 1_{\eye} \otimes f VV  @VV f V\\
\eye \otimes A' @>>\lambda_{A'}> A'
\end{CD}
\hspace{2cm}
\begin{CD}
A \otimes \eye @>\rho_{A}>> A\\
@V f \otimes 1_{\eye} VV  @VV f V\\
A' \otimes \eye @>>\rho_{A'}> A'
\end{CD}
\]
\end{enumerate}
In order to extend the associativity and proper unit object interaction to
all possible $n$-ary tensor products,
the natural isomorphisms must fulfill two coherence axioms, \ie the
following diagrams must always commute:
\begin{enumerate}
\item[(i)] (Pentagon axiom)
\[
\xymatrixcolsep{5em}
\xymatrix{
((A \otimes B) \otimes C) \otimes D \ar[r]^{\alpha_{A \otimes B,C,D}} \ar[d]_{\alpha_{A,B,C} \otimes 1_{D}} &
(A \otimes B) \otimes (C \otimes D) \ar[r]^{\alpha_{A,B, C \otimes D}} &
A \otimes (B \otimes (C \otimes D)) \\
(A \otimes (B \otimes C)) \otimes D \ar[rr]_{\alpha_{A, B \otimes C, D}} & &
A \otimes ((B \otimes C) \otimes D) \ar[u]_{1_{A} \otimes \alpha_{B,C,D}}
}
\]
\item[(ii)] (Triangle axiom)
\[
\xymatrixcolsep{2em}
\xymatrix{
(A \otimes \eye) \otimes B \ar[rr]^{\alpha_{A,\eye,B}} \ar[dr]_{\rho_{A} \otimes 1_{B}} & &
A \otimes (\eye \otimes B) \ar[dl]^{1_{A} \otimes \lambda_{B}} \\
& A \otimes B
}
\]
\end{enumerate}
\end{definition}

\begin{remark}[Strict monoidal categories]
If all the natural transformations are identities we have
$(A \otimes B) \otimes C = A \otimes (B \otimes C)$ and
$A \otimes \eye = A = \eye \otimes A$ for all objects, and
$\cat{C}$ is said to be \emph{strict}.
However, this is not as important as it might seem;
Every monoidal category is monoidally equivalent to a strict
monoidal category --- this is known as Mac Lane's strictification
theorem~\cite{MacLane}.   Intuitively this says that we don't lose much by
considering equalities instead of isomorphisms.
\end{remark}

\begin{definition}[Symmetric monoidal category]
Adding a further family of natural isomorphisms called
\emph{symmetric braidings},
$c_{A,B}: A \otimes B \to B \otimes A$
with the property $c_{B,A} = c_{A,B}^{-1}$ 
fulfilling the
\begin{enumerate}
\item[(i)] (Hexagon axiom)\[
\begin{CD}
(A \otimes B) \otimes C  @> c_{A,B} \otimes 1_C >>
(B \otimes A) \otimes C  @>\alpha_{B,A,C}>> B \otimes (A \otimes C)\\
@V \alpha_{A,B,C} VV  @.  @VV 1_B \otimes c_{A,C} V\\
A \otimes (B \otimes C) @>>c_{A, B \otimes C}>
(B \otimes C) \otimes A @>>\alpha_{B, C, A}>
B \otimes (C \otimes A)
\end{CD}
\]
\end{enumerate}
makes a monoidal category \emph{symmetric}.
Intuitively the symmetric braidings mean that the relative order of
the objects in a tensor product carries no fundamental significance.
\end{definition}

\begin{definition}[Compact closed category]
A \emph{compact closed} category $\cat{C}$ is a symmetric monoidal category
in which for every object $A \in \Ob{C}$ there is a
\emph{dual object} $A^* \in \Ob{C}$, 
and \emph{unit} and \emph{counit} morphisms
$\eta_A: \eye \to A^* \otimes A$ and
$\epsilon_A: A \otimes A^* \to \eye$, which fulfill the adjunction
triangle equations
\begin{align}
\notag
(\epsilon_A \otimes 1_A) \circ (1_A \otimes \eta_A) &= 1_A,\\
(1_{A^*} \otimes \epsilon_A) \circ (\eta_A \otimes 1_{A^*}) &= 1_{A^*}.
\label{eq:adjtriangle}
\end{align}
In quantum mechanics, the unit and counit morphisms can be thought of as
corresponding to generalized Bell states.
\end{definition}

\begin{definition}[Dagger-compact category]
A \emph{dagger}-compact category is a compact closed category that 
comes equipped with a contravariant \emph{dagger} endofunctor
$\dagger: \cat{C} \to \cat{C}$,
which behaves precisely in the way one would expect from the
Hermitian adjoint operation on a Hilbert space~\cite{catQM}.
To make the analogy even stronger, an isomorphism $f$ is said
to be \emph{unitary} iff $f^\dagger = f^{-1}$.

The dagger functor has the following properties:
\begin{enumerate}
\item[(1)]
Identity on the objects: $A^\dagger = A$.

\item[(2)]
Associates every morphism $f: A \to B$ with its
\emph{adjoint} morphism $f^\dagger: B \to A$.

\item[(3)]
Involutivity: $f^{\dagger \dagger} = f$, and

\item[(4)]
Compatibility with the tensor product:
$(f \otimes g)^\dagger = f^\dagger \otimes g^\dagger$.
\end{enumerate}
Furthermore, the natural isomorphisms $\alpha$, $\lambda$, $\rho$ and $c$ must
all be unitary, and the unit and counit morphisms must fulfill
$\eta_A = c_{A,A^*} \circ \epsilon_A^\dagger$.
\end{definition}

\begin{remark}[Scalars in a symmetric monoidal category]
Scalars in a monoidal
category are maps from the tensor unit $\eye$ back to the tensor unit
$\eye$.  We consider some (say) complex number $s$, this number is then a
map of type $s:\eye\to\eye$.  By linearity, $s(1)$ completely
determines the map, and hence the scalar $s$.   We note that the scalars
in a SMC form a commutative monoid~\cite{KL80}.  
\end{remark}


\section{Uniqueness of the cup state}

\begin{figure}[h!]
\includegraphics[width=0.4\textwidth]{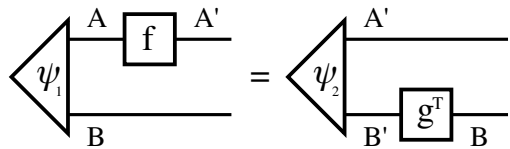}
\caption{``Sliding'' linear operators around a cup state.}
\label{fig:cup-proof}
\end{figure}

Assume that given the Hilbert spaces $A$ and $A'$, we can write down two bipartite states
\begin{align}
\ket{\psi_1^A} &= c_{xy} \ket{x}_A \ket{y}_B,\\
\ket{\psi_2^{A'}} &= d_{xy} \ket{x}_{A'} \ket{y}_{B'}
\end{align}
that should play the
role of the cup, where the complex coefficients $c_{xy}$ and $d_{xy}$ can be
interpreted as elements of the matrices~$C$ and~$D$. The normalization condition gives
\begin{align}
\braket{\psi_1}{\psi_1} = c^*_{xy} c_{xy} = \Tr(C^\dagger C) = 1,\\
\braket{\psi_2}{\psi_2} = d^*_{xy} d_{xy} = \Tr(D^\dagger D) = 1.
\end{align}
We require that the pair of states should have the following property:
For every linear operator
\[
f: A \to A', \quad f = f_{ij} \ket{i}_{A'} \bra{j}_A
\]
there is another linear operator
\[
g: B \to B', \quad g = g_{ij} \ket{i}_{B'} \bra{j}_B
\]
and vice versa such that the graphical equality
in~\Figref{fig:cup-proof} holds;
we want to be able to ``slide'' the operators around the cup.
For the sliding operation to make sense, the dimensions of the external legs
must remain the same.

In equation form this is
\begin{align}
\notag
& f_{ij} c_{xy} \ket{i}_{A'} \braket{j}{x}_A \ket{y}_B
= f_{ij} c_{jy} \ket{i}_{A'} \ket{y}_B
= f_{ij} c_{jk} \ket{i}_{A'} \ket{k}_B\\
&=
(g^T)_{ij} d_{xy} \ket{x}_{A'} \braket{j}{y}_{B'} \ket{i}_{B} 
= d_{xj} g_{ji} \ket{x}_{A'} \ket{i}_{B}
= d_{ij} g_{jk} \ket{i}_{A'} \ket{k}_{B},
\end{align}
which is equivalent to the matrix equation
\begin{align}
\label{eq:cup-cond}
F C = D G.
\end{align}
Now let us apply the singular value decomposition
on $C$ and~$D$.
SVD is given by $X = U \Sigma V^\dagger$,
where $U$ and $V^\dagger$ are unitary and $\Sigma$ is a
diagonal matrix with the (real, nonnegative) singular values
$\sigma_k$ of $X$ on the diagonal in nonincreasing order.
We obtain
\begin{align}
\notag
& F (U_C \Sigma_C V_C^\dagger) = (U_D \Sigma_D V_D^\dagger) G\\
\notag
\Leftrightarrow \quad &
(U_D^\dagger F U_C) \Sigma_C  = \Sigma_D (V_D^\dagger G V_C)\\
\Leftrightarrow \quad &
\tilde{F} \Sigma_C  = \Sigma_D \tilde{G}.
\end{align}
The state normalization condition is equivalent to $\sum_k \sigma_k^2 = 1$.
Thus for both $C$ and $D$ we always have $\sigma_1 > 0$.
Comparing the elements $(\dim A',1)$ and $(1,\dim B)$ on both sides
we find that unless $\dim A \ge \dim B$, $\dim A' \le \dim B'$,
${\sigma_C}_{\dim B} > 0$ and ${\sigma_D}_{\dim A'} > 0$,
there are either matrices $F$ for which there is no $G$ such that 
\Eqref{eq:cup-cond} holds, or vice versa.

Now assume we wish to impose two additional constraints:
The cup states have to be symmetric and
map unitary operators to unitary operators.
The first constraint gives $A \isom B$ and $A' \isom B'$,
which means that $\Sigma_C$ and $\Sigma_D$ are both square and, since
their singular values are all positive, invertible:
\begin{align}
\tilde{G} = \Sigma_D^{-1} \tilde{F} \Sigma_C
\quad \Leftrightarrow \quad
G = D^{-1} F C.
\end{align}
Furthermore, for every unitary matrix $\tilde{F}$ we
must have
\begin{align}
\tilde{G} \tilde{G}^{\dagger} 
= \Sigma_D^{-1} \tilde{F} \Sigma_C \Sigma_C^\dagger \tilde{F}^\dagger (\Sigma_D^{-1})^\dagger 
= \Sigma_D^{-1} \tilde{F} \Sigma_C^2 \tilde{F}^\dagger \Sigma_D^{-1}
= \I
\quad \Leftrightarrow \quad \tilde{F} \Sigma_C^2 = \Sigma_D^2 \tilde{F}.
\end{align}
Now Schur's Lemma for unitary representations of Lie groups says that
$\Sigma_C$ and $\Sigma_D$ must both be scalar multiples of identity.
Together with the normalization condition this 
completely fixes the singular values, and we can choose
$C = \frac{1}{\sqrt{d_A}} U$ and
$D = \frac{1}{\sqrt{d_{A'}}} V$, where $U$ and $V$ are symmetric unitary matrices.
Thus
\be
\ket{\psi_1^A} = c_{xy} \ket{x}_A \ket{y}_A
= \frac{1}{\sqrt{d_A}} {U}_{xy} \ket{x}_A \ket{y}_A
= (U \otimes \I) \cupket{A},
\ee
and the most general cup state for the Hilbert space $A \otimes A$ is a local unitary rotation of~$\cupket{A}$.

\bibliographystyle{apsrev4-1}
\bibliography{qc}

\end{document}